\documentclass{article}

\PassOptionsToPackage{numbers,sort&compress}{natbib}
\usepackage[preprint]{neurips_2026_vericode}

\usepackage{booktabs}                    
\usepackage{tikz}                        
\usepackage{float}                       
\usepackage{array}                       
\newcolumntype{L}[1]{>{\raggedright\arraybackslash}p{#1}}
\usepackage{xcolor}                      
\usepackage{enumitem}                    
\usepackage{url}                         
\providecommand{\doi}[1]{doi:\discretionary{}{}{}\texttt{#1}}
\DeclareFontFamily{OMS}{cmtt}{\skewchar\font=48}
\DeclareFontShape{OMS}{cmtt}{m}{n}{<->ssub*cmsy/m/n}{}

\title{A Trust Ledger and an Execution Check for CPG-Based C-to-Lean 4 Autoformalization:\\
Separating Declined from\\Silently Incorrect Translations}

\author{%
  Ishan K Singavarapu\\
  Independent\\
  \texttt{ishansingavarapu@gmail.com}
  \And
  Manish Bhatt\\
  OWASP\\
  \texttt{manishbhatt13212@gmail.com}
}

\begin{document}
\maketitle

\begin{abstract}
Verifying large C codebases requires translating them into formally checkable semantics, but most autoformalization work reports a single aggregate success rate. That rate conflates two different failures: code the translator declined to handle, and code it translated incorrectly. We present a deterministic code-property-graph (CPG)-based exporter that translates C source into a small Lean~4 core semantics and reports three structural coverage measures: hole-free, call-closed, and dynamic-hole risk. They keep what was declined apart from what was attempted, but none of the three is a correctness claim, and none can tell a correct translation from a silently incorrect one without executing it. Applied to a CPG export of SQLite (8{,}602 functions, over 1.5~million AST nodes), 5{,}222 functions (60.7\%) are hole-free, but only 2{,}134 (24.8\% of the export) are call-closed. The hole-free population is $2.45\times$ the call-closed one, which is a gap that a single rate wouldn't reveal. Of six cases where coverage and fidelity moved independently, three are silent wrong answers: a character literal, a do-while loop, and a variadic signature each translated with no hole and computed the wrong result, and no coverage number registered the defect or its repair. In another, we retracted our own ``impossible'' classification of function-pointer dispatch after tracing a counterexample in SQLite, which changed 17 translations and moved no number. Our main contribution is methodological: a protocol that keeps declined and silently incorrect translations apart, and separates constructs no analysis inside the program can resolve from ones that only look that way. A silent wrong answer is more dangerous than any hole, because a hole is counted and a wrong answer isn't.
\end{abstract}

\section{Introduction}

An autoformalization pipeline turns code into an artifact a proof assistant can check, and
reports how much of the input it managed. That number carries an implicit promise: raising
it means the translation got better. The promise is rarely stated, because it sounds
like a truism: translating more of a program must mean translating it better.

This paper separates these two distinct claims. We report six cases of decoupling from a
deterministic code-property-graph (CPG) based pipeline that translates C into a small Lean~4
interpreter, with one finding about the metric's own vocabulary. Four of the six are
correctness defects whose repair moved no coverage number, or in A4's case barely moved one
(A1--A4). One is a coverage movement that was mostly a relabeling (B1). One is a corrected
impossibility claim that changed 17 translations and moved no coverage number either (C1). The
seventh finding, T1, is a defect in the cause taxonomy that the coverage report is built from. It
is not a case of independent movement, which is why six cases yield these seven findings.

Table~\ref{tab:findings} lists all seven under these labels. Each is a dated change to the
pipeline, with the values it was checked against.

The distinction we present is between two ways that a translation can fail. A construct the
translator \emph{declined} to handle leaves an explicit and labeled gap, and coverage can count it. A construct the translator handled \emph{incorrectly} produces well-typed output that
type-checks, is counted as translated, but computes the wrong answer. By construction, coverage cannot see
the second class: it is a fold over the artifact and the wrong answer is a
property of running the artifact. That much is an observation about definitions. What this
paper adds is that the gap is not hypothetical: it is populated by ordinary constructs such as
character literals, do-while loops, and variadic signatures. It also runs in both directions, so
movement in the metric is not evidence either way.

Two terms recur throughout: a function is \emph{hole-free} when nothing in its translated syntax tree is a placeholder standing in for a construct the translator declined, and \emph{call-closed} when it is additionally free of direct calls that do not resolve to an
exported function. \S2 states both precisely.

\paragraph{Three outcomes.} A construct ends up in one of three states: declined, translated
correctly, or translated incorrectly with no signal. The ledger separates \emph{declined} from
\emph{attempted}, per construct rather than in aggregate. It cannot separate \emph{correctly
translated} from \emph{silently incorrect}, because that distinction is not a property of the
artifact that the ledger folds over. Recovering it requires executing the translated program
against a reference implementation. The discipline reported here is therefore the three
structural measures and an execution check, and \S3.1 is the argument for the second half. We
reserve ``verified'' for properties a proof assistant has checked, and none of the three measures
is one.

\paragraph{Why existence claims.} Our claims are existential. To
show that coverage and fidelity are decoupled, one case in each direction suffices, and a
case is not weakened by having been observed on a single pipeline. We do not make a claim about
how \emph{often} they decouple, or that the magnitudes here transfer. Where we report a
magnitude, such as the $2.45\times$ denominator effect in \S4, it is a measurement on one
codebase and labeled as such. Thus, the paper gives a protocol and instances, not a rate.

\paragraph{What has been checked against an oracle.} For SQLite, no hole-free or call-closed
function has been checked against an oracle at population level. The executed checks are
hand-built fixtures (\ref{app:fixtures}). \S5 says what a population-level check would require.

\paragraph{Contributions.}
\begin{enumerate}[leftmargin=1.5em,itemsep=2pt,topsep=3pt]
\item \textbf{An evaluation protocol for separating declined from silently incorrect} (\S3): the
methodological contribution. The ledger counts what was declined, by cause, and an execution check
against the language implementation catches what was attempted and answered wrongly. Applied to
one pipeline, the two together produced six cases: four correctness defects repaired with little
or no metric movement, one metric movement that was mostly a relabeling, and one retracted
impossibility claim that moved the metric by nothing while changing 17 translations. The
retraction separates constructs no analysis inside the program can resolve, such as a public API
entry point, from those that only look that way until someone traces one (\S3.3).

\item \textbf{Instability of the cause taxonomy} (\S3.4): the number of distinct hole causes is not a stable quantity, because a label built from source text partitions on the source.

\item \textbf{The Denominator Effect} (\S4): if read purely as coverage, the reported figure on
SQLite differs by $2.45\times$ depending on whether call closure is required, and containment
makes some gap certain while its size is what's measured.
\item \textbf{What we report and its limits} (\S5): three non-collapsible tiers,
holes itemized by cause, and an execution check. The first two are not enough on their own,
since \S3.1 is a list of defects that neither the tiers nor the itemization caught, and the third
has so far run only on fixtures.
\end{enumerate}

\section{The pipeline and the metric}

The pipeline is context for the findings, not a contribution, so we only describe what is needed to read \S3 and \S4.

A definitional interpreter for a compact universal source language (``Core'') is written
once in Lean~4 \citep{demoura2021}. It is fuel-indexed and total, structurally recursive on
a fuel budget, with no \texttt{partial} and no \texttt{sorry}. A codebase is mechanically
transpiled into a term of that interpreter's syntax type, so a program becomes an ordinary
Lean data value and every property of it is a theorem about \texttt{eval} applied to an AST. Because the semantics is a computable interpreter, the translated program can
be run and compared against the language implementation, which is what \S3.1 relies on.

Joern's code property graph \citep{joern,yamaguchi2014} supplies a single node vocabulary, so the
pipeline writes one CPG-to-JSON exporter and one JSON-to-Lean printer, both deterministic,
with no model on the translation path. Two consequences matter here. First, a construct the
exporter cannot translate faithfully becomes a \emph{hole} tagged with the CPG label that
produced it, so declined constructs are counted by cause rather than in aggregate. Second,
the whole system is downstream of one parser version, pinned in the artifact
(\texttt{joern-version} at 4.0.606, \texttt{lean-toolchain} at
\texttt{leanprover/lean4:v4.30.0-rc1}).

\paragraph{Defining the metric.} The ledger reports three structural coverage measures and never
collapses them into one score. Table~\ref{tab:defs} gives the unit of each, along with what it does
and does not establish. None of the three is a correctness claim. Call closure is the one most
easily over-read. In the AST, a call to a function that was never translated looks the same as a
call to one that was, so this tier removes that failure mode, but only for \emph{direct} call
sites. It doesn't follow calls transitively, and says nothing about what a resolved callee does.
\emph{Dynamic-hole risk} counts constructs that may still hole on some input. We report it for
completeness, but the findings below turn on the first two tiers and do not depend on it. \S3 is
about what these numbers do and do not track.

\begin{table}[t]
\centering\scriptsize\setlength{\tabcolsep}{3.5pt}
\caption{The terms the paper uses, with the unit each is counted in. Percentages are over exported
functions. The last column is what the term doesn't establish.}
\label{tab:defs}
\begin{tabular}{L{0.12\linewidth}L{0.08\linewidth}L{0.45\linewidth}L{0.28\linewidth}}
\toprule
Term & Unit & Definition & Does not establish \\
\midrule
exported & function & A function the Joern CPG export contains and the printer attempted. It is the denominator of every tier. A definition the parser drops never enters it. &
  that the function was translated, or that the export covers the source tree \\
hole & occurrence & A placeholder AST node, tagged with the CPG label that produced it, standing in
  for a construct the translator declined. & --- \\
hole-free & function & An exported function whose translated AST contains no hole. &
  that the function computes what the C does \\
call-closed & function & Hole-free, and every direct call site names a function that resolves
  inside the export, by an exact name match or a unique dotted-suffix match. No library name is
  assumed known for C. It is not transitive, so a callee's own holes are not examined, and a
  self-call resolves. &
  that calls through a computed callee, such as a function pointer, callback, or vtable, resolve,
  since they have no name and are not checked, or what any callee does \\
dynamic-hole risk & construct & A construct that may hole on some input, such as a field access,
  an index, a method call, or arithmetic that can hit undefined behavior. Counted, not subtracted. &
  whether it does hole, which depends on the input \\
\bottomrule
\end{tabular}
\end{table}

\paragraph{One structural blind spot, stated up front.} Joern parses C without running the
preprocessor, so a definition whose signature defeats the parser is not holed but \emph{absent},
with no row anywhere in the ledger. Therefore, coverage is an accounting of what was attempted, not
of what the codebase contains. This is separate from the decoupling result, so we don't count it
among the seven findings.

\section{Coverage and fidelity move independently}

This section gives the six cases and the one finding about vocabulary, grouped by which way the
translation and the metric moved. Four go one way, one goes the other, and one is a claim with
zero effect on the metric.

\begin{table}[t]
\centering\scriptsize\setlength{\tabcolsep}{3.5pt}
\caption{The findings: what each change did to the translated program (behavior) beside what it
did to the coverage metric. The second column reads behavior\,/\,metric. Rows~A are correctness
changes the metric did not register (A4 barely did), Row~B is a metric movement that was mostly a
relabeling, Row~C is a corrected impossibility claim, and Row~T is a defect in the metric's
vocabulary, so it has no behavior\,/\,metric entry. Corpora and commit labels are given per case
in the text and in \ref{app:prov}.}
\label{tab:findings}
\begin{tabular}{L{0.032\linewidth}L{0.075\linewidth}L{0.26\linewidth}L{0.215\linewidth}L{0.15\linewidth}L{0.135\linewidth}}
\toprule
& behavior / metric & Finding & What changed & What the metric did & How it was found \\
\midrule
A1 & YES / no & Character literal rendered as a string, so \texttt{*p == 'x'} compared an integer with a string
   & the loop returned 0 for every input, and after the fix 3, 2, 0, matching compiled \texttt{cc}
   & the hole count stayed at 5,033 and no label moved
   & executing the generated code \\
A2 & YES / no & \texttt{do B while(C)} emitted as \texttt{while(C) B}
   & \texttt{cc -O0} returns 1 and the translation returned 0, and 7 functions changed
   & both shapes are hole-free, so no coverage number separates them
   & a hand-built fixture against \texttt{cc} \\
A3 & YES / no & C variadic signatures marked as Python varargs
   & surplus arguments packed as a tuple under the wrong calling convention, for every printf-style function
   & no hole emitted, before or after
   & a test suite meeting a large corpus \\
A4 & YES / barely & Struct local whose address escapes to an in-program call
   & statically hole-free, but failed at evaluation with \texttt{setField:*:non-object}
   & hole-free $+0.05$ and $+0.12$ points on two scopes
   & a differential-test fixture \\
\midrule
B1 & little / YES & Sentinel casts had been mislabeled
   & 16 of 31 moved to a more accurate cause, 15 closed
   & one label $-31$, another $+16$, and hole-free $+1$
   & reading the labels, not the total \\
\multicolumn{6}{p{0.95\linewidth}}{\emph{\,\,second instance of B1: interior-pointer labels reclassified to a more precise cause, one label 10 to 5, none of the five translatable}} \\
\midrule
C1 & YES / no & Vtable dispatch proven closed after being recorded as impossible
   & the translated output of 17 functions changed, and 19 (function, parameter) pairs closed
   & zero net new hole-free functions
   & a full content diff \\
\midrule
T1 & --- & Hole labels built from source text
   & every source fragment became its own ``cause''
   & 15 distinct causes collapsed to 2
   & a label-taxonomy test \\
\bottomrule
\end{tabular}

\end{table}

\subsection{Correctness changed and the metric did not move, or barely moved}

\paragraph{A1: a character literal rendered as a string.} The exporter's literal dispatch had no
case for a bare single-quoted literal, so it fell through to a generic ``starts with a quote''
branch and rendered \texttt{'x'} the same way as \texttt{"x"}. In C, a single-quoted literal is a
number: the character's own codepoint. Reading a byte through a tracked cursor already produced
that byte's integer value correctly, so \texttt{*p == 'x'} compiled into a comparison of an integer
with a string, which the semantics' equality never treats as equal. Therefore, a byte-counting loop
compiled with zero holes and returned 0 for every input. The repair recognizes numeric character
literals in the pipeline's C-like \emph{dialect}, the front-end grammar it uses for C, as distinct
from its Python and JavaScript ones. It handles the standard escapes and leaves a multi-character
literal or an unrecognized escape to fall through unchanged, instead of guessing. We checked it by
execution: over \texttt{"axxbxcdef"} at $n = 9, 3, 1$, the loop now returns 3, 2, 0, which matches
compiled \texttt{cc}.

On the bounded SQLite corpus of 3,560 functions, the hole count and \emph{every individual label}
were unchanged, at 5,033 before the fix and 5,033 after. That corpus, the denominator for this
case, is an iteration corpus smaller than the 8,602-function export of \S4. Every function
comparing a byte against a letter literal became correct, but no number in the coverage report
recorded it. Digit literals stayed wrong for a separate reason we found later: \texttt{'0'} parsed
as 0, not 48. This is the cleanest case in the paper, because the whole corpus was measured before
and after, and the metric didn't move even slightly. Tracked by coverage alone, the commit would
have looked like a no-op.

Like A2 and A3 below, this case is pinned by a committed specification (\ref{app:fixtures}). It
builds the same loop twice inside the semantics, and the two versions differ only in the literal
the byte is compared against. We keep the defective version as a negative control, because the
defect never appeared as a hole and nothing in the ledger would notice a regression to it.

\paragraph{A2: a do-while loop emitted as a while loop.} The exporter matched \texttt{case "WHILE"
| "DO"} and emitted the same loop statement for both. A do-while runs its body at least once, while
the emitted form tests first, so the two disagree whenever the condition starts out false. Checked
against \texttt{cc -O0}, \texttt{do \{ n = n + 1; \} while (0)} returns 1 and the translated while
form returned 0. The repair emits \texttt{while (true) \{ B; if (C) skip else break \}} and
rewrites each \texttt{continue} that belongs to that loop, because a do-while's \texttt{continue}
jumps to the condition test and the \texttt{while(true)} shape has no test to reach. Seven
functions in a C++ corpus changed.

No hole was involved at any point. The construct translated to a statement before the fix and to a
statement after it, so both versions are hole-free and no coverage number can tell a correct loop
from an incorrect one.

The case rests on three links, each checkable on its own: the compiler disagreement, the line
\texttt{case "WHILE" | "DO"} that the diff splits in two, and a specification that keeps the
defective shape as a negative control (\ref{app:fixtures}). If the metric can't tell the two shapes
apart, the test suite has to.

\paragraph{A3: C variadic signatures treated as Python varargs.} The exporter gated one
Python-specific parameter analysis on the file extension, but it did not gate the variadic flag,
which C and C++ set for \texttt{...}. Therefore, a signature such as \texttt{void V8\_Fatal(char*,
...)} carried a vararg marker, which told parameter binding to pack surplus arguments into a tuple
under Python's calling convention. Every printf-style function in the corpus was silently
mistranslated. C's \texttt{...} is read through \texttt{va\_arg}, which the pipeline does not
translate, so the marker is now not emitted for non-Python sources at all. No hole was emitted
before or after, so the metric was blind to both the defect and its repair.

This case is pinned like A1 and A2. What it adds is the paper's distinction in its most direct
form. A committed specification builds the same signature twice, differing in one field. Under the
marker, a three-argument call silently binds the two surplus arguments as a tuple. With the marker
removed, the same call is refused. Neither one models \texttt{va\_arg}. That is the point of the
case, not an omission: the defect turned a call the pipeline would have \emph{declined} into one it
\emph{answered wrongly}. Both values are pinned, and the defective shape is kept as a negative
control.

\paragraph{A4: a struct address escaping to an in-program call.} A struct local whose address is
passed to another translated function was left out of the boxing analysis, while field access was
emitted unconditionally for any class-typed receiver. The result translated as statically hole-free
and then failed at evaluation with \texttt{setField:*:non-object}. We confirmed that the defect
predates the work that found it by checking it against an earlier merge commit. The repair moved
hole-free by 0.05 and 0.12 percentage points on two SQLite scopes. That is movement, but it is the
size of a rounding error, not of a function that now returns the right value.

\subsection{The metric moved more than the translation did}

\paragraph{B1: an apparent 31-hole improvement that was mostly a relabeling.} One change, measured
on a 3,807-function SQLite iteration corpus, reduced \texttt{op:cast:opaque-type} from 46 to 15
occurrences and raised \texttt{op:cast:pointer:int-to-pointer} from 14 to 30. Most of that wasn't
a closure. Sentinel casts of the \texttt{SQLITE\_TRANSIENT} family had been mislabeled as an opaque
type, and 16 of the 31 occurrences now carry the right cause: a pointer cast whose operand is not
pointer-shaped, which is still a hole. The other 15 did close, through the same commit's handling
of enum casts and function-pointer typedef targets, and hole-free moved by one, across 14 changed
functions. Someone tracking that label would record a 31-occurrence improvement, while someone
tracking hole-free would record one function. Thus, most of what got more accurate was the
pipeline's \emph{description} of a construct, not its treatment of it.

An earlier commit in the same family is a smaller instance, which we count with it.
\texttt{op:addressOf:element:opaque-type} fell from 10 to 5 because five occurrences took the more
precise \texttt{op:addressOf:element:object}, and none of them became translatable.

\subsection{A corrected claim, with no movement either way}

\paragraph{C1: a retraction that moved nothing and changed 17 functions.} Function-pointer and
vtable dispatch had been recorded as a \emph{fundamental} limit of whole-program static analysis,
in the same category as a public API entry point called from outside the analyzed program. Once we
traced the claim instead of defending it, it didn't survive. SQLite assigns its own function
pointers to its own functions and calls them entirely within the analyzed source, such as
\texttt{pPage->xCellSize = cellSizePtr;} and \texttt{pPage->xCellSize(pPage, \&data[pc])} in
\texttt{btree.c}. The exporter had been \emph{choosing} not to prove closure, through a blanket
``address ever taken, therefore unclosable'' guard, and not for lack of information. Widening an
existing scan to record plain function-pointer field assignments closed 19 (function, parameter)
pairs.

The yield in coverage was zero net new hole-free functions on the 3,807-function SQLite iteration
corpus. Part of that is by definition, since call closure doesn't examine computed callees
(Table~\ref{tab:defs}), and the functions involved were already hole-free. Even so, a full content diff found 17
functions whose translated output changed while their hole count stayed at zero. Targeted
elaboration confirmed that the new output is well-typed. Thus, this one change shows the decoupling
in both directions at once: a wrong classification was corrected and 17 translations changed, but
no tier of the metric registered any of it.

Separately, the same change caught a defect of its own, which is the same phenomenon one level
down. The new whole-program fixed point depended on a table that only fills as the priming passes
run. Declared as a \texttt{lazy val}, it memoized an empty snapshot on first access and never
recomputed, so the mechanism measured zero effect until we made it a \texttt{def}. A check that shares an assumption with the thing it checks will
flatter it, and here the shared assumption was a cache.

\paragraph{Three classes of limit, and what C1 does and does not retract.} C1 shows that
``unresolvable'' was filed under one heading when the cases differ in what would resolve them. We
separate three classes, drawn from the constructs in this paper. The first is
\emph{boundary-dependent}: the missing fact lives outside the analyzed program, as for the caller
of a public API entry point. No analysis inside the program recovers it, but a specification of the
boundary may, so we do not call this limit fundamental. The second is \emph{analysis-recoverable}:
the program contains the fact and the exporter declined to derive it, as for function-pointer
dispatch in C1, which more whole-program analysis closed. The third is
\emph{representation-limited}: the construct has no faithful translation into the Core's value
semantics, as for much of the pointer and address arithmetic in \S4, so more analysis does not help
and only a change of semantics could. C1 moves one construct family from the first class to the
second. It doesn't show that the first or third class is empty, and we haven't audited the rest of
the ledger for other misfiled categories.

\subsection{The metric's vocabulary is not stable either}

\paragraph{T1: labels built from source text.} The remedy for an aggregate rate is to itemize
declined constructs by cause (\S5). That presumes the set of causes is a property of the pipeline.
It wasn't. The \texttt{stmt:UNKNOWN:} label carried the first word of the offending source, so one
C++ corpus alone produced separate ``causes'' named \texttt{)}, \texttt{\}}, \texttt{V8\_WEAK;},
and a fragment of a C++ \texttt{requires} clause. Because the ledger groups by label, every source
fragment became its own cause. Keying the label on the parser node type instead collapsed 15
distinct labels to 2, both of which name a remedy, not a symptom.

The consequence is narrow. The ``number of distinct hole causes'' can't be compared across
pipelines, or even across corpora of the same pipeline, without knowing how the labels are built.
We report 70 causes in \S4 for one specific export, not as a property of the approach.

\subsection{What surfaced them, and what did not}

None of the four correctness defects (A1--A4) was surfaced by the coverage rate or by the itemized
inventory (Table~\ref{tab:findings}). Only the two B1 relabelings were visible in the breakdown,
and both are obvious once labels are read per cause. That is the argument for itemizing rather than
reporting a rate.

\section{Read purely as coverage, the reported figure depends on the denominator}

\S3 argues that coverage does not track fidelity. Coverage is still a useful report of what was
attempted. This section measures how much the reported number depends on a choice that is easy to
leave implicit. All figures come from one committed export of SQLite \citep{sqlite}
(\texttt{autoform\_sqlite\_numbers.json}, module \texttt{Sqlite}, dialect \texttt{c-like}),
summarized in Figure~\ref{fig:tiers}. It contains 8,602
functions and 1,513,464 AST nodes, which is what the export saw and not a count of the SQLite
source tree (\S2). Of those functions, 5,222 (60.7\%) are hole-free and 2,134 (24.8\%) are
call-closed. Its 14,260 residual holes are itemized under 70 causes. Dynamic-hole risk is counted
per construct, not per function, and 216,730 constructs carry it. As \S3.4 shows, the 70-cause
figure is a property of the labeling scheme, not of the codebase, though here the labels are keyed
on node type and none of the 70 is a source fragment.

\begin{figure}[H]
\centering
\begin{tikzpicture}[every node/.style={font=\footnotesize,inner sep=0pt}]
  \draw[fill=black!3,draw=black!45] (0,0) rectangle (10.4,4.10);
  \node[anchor=north west] at (0.12,4.02)
    {\textbf{exported} 8,602 functions, 1,513,464 AST nodes};

  \draw[fill=black!10,draw=black!55] (0.35,1.70) rectangle (6.4,3.55);
  \node[anchor=north west] at (0.47,3.47) {\textbf{hole-free} 5,222 (60.7\%)};

  \draw[fill=black!28,draw=black!75] (0.75,2.05) rectangle (3.2,3.00);
  \node[anchor=north west,text=white] at (0.87,2.92) {\textbf{call-closed}};
  \node[anchor=north west,text=white] at (0.87,2.55) {2,134 (24.8\%)};

  \node[anchor=north west,font=\scriptsize,text width=2.95cm] at (3.34,3.00)
    {hole-free is $2.45\times$ call-closed, and the rest have an unresolved callee};

  \node[anchor=north west] at (6.62,3.47) {\textbf{declined}, itemized};
  \node[anchor=north west,font=\scriptsize] at (6.62,3.14) {14,260 holes / 70 causes};
  \node[anchor=north west] at (6.62,2.64) {\textbf{dynamic-hole risk}};
  \node[anchor=north west,font=\scriptsize] at (6.62,2.31) {216,730 constructs};

  \draw[dashed,thick,draw=black!80] (1.3,0.42) rectangle (5.6,2.20);
  \node[anchor=north,align=center,font=\scriptsize,text width=4.15cm] at (3.45,1.58)
    {\textbf{silently incorrect}\\
     well-typed, counted as translated,\\
     but computes the wrong answer\\
     no tier bounds it, and A1--A3 are here};
\end{tikzpicture}
\caption{Call-closed functions are a subset of hole-free functions, which are a subset of exported functions, so a gap of some size is certain and only the magnitude is our measurement. The dashed region is the class this paper is about. It crosses every
tier boundary because it is not a property of the artifact the tiers fold over: a silently
incorrect translation is hole-free, may be call-closed, and is counted as translated. No
denominator excludes it and no cause itemizes it, which is why recovering it needs an execution
check and not a better rate. Areas are not to scale. Only the three sets on the left are nested,
and the right-hand column counts occurrences and constructs, not functions.}
\label{fig:tiers}
\end{figure}

Call-closed is a subset of hole-free by construction, so a gap is guaranteed and isn't a result in
itself. What is measured is its size: $2.45\times$. The functions in between have no holes of their
own but make a direct call that does not resolve inside the export, so their behavior is unknown
for a reason no hole count inspects. No library name is assumed known for C, so those calls include
every call into the C library, and we haven't split the gap by callee. Therefore, reporting hole-free alone overstates the
population with no unresolved direct call, by that factor on this codebase. We do not claim the
factor transfers. What we claim is that the choice of denominator is a reporting decision with a
measurable effect, and that a single coverage rate doesn't say which choice was made.

The distribution of declined constructs is informative in its own right. It is dominated by C
pointer and address semantics, which are precisely the constructs that look like ordinary
operations but admit no faithful value-semantics translation. The ten largest causes are
\texttt{cstr:address-compare} (1,415), \texttt{cstr:pointer-arith} (1,338),
\texttt{assign:lhs:indirection} (1,150), \texttt{op:addressOf:element:scalar} (1,121),
\texttt{op:cast:pointer:int-to-pointer} (952), \texttt{op:indirection:pointer} (947),
\texttt{op:cast:opaque-type} (720), \texttt{control:GOTO} (503), \texttt{op:postIncrement:pointer}
(492), and \texttt{stmt:UNKNOWN:CASTProblemDeclaration} (463). Another 5,159 occurrences are spread
across 60 further causes. Each is a declined construct with a named boundary, not a silent
mistranslation, which is the property \S3 shows coverage does not otherwise guarantee.

\section{What we report instead, and why it is not sufficient}

The constructive proposal is modest, and \S3 is the argument for its insufficiency, not its
adequacy. It comes down to three things, in order of cost. First, report three tiers and never
collapse them, so that the denominator is explicit. \S4 shows that choice is worth $2.45\times$ on
this corpus, though it's usually left implicit. Second, itemize every declined construct by its
cause, so that ``the other 40\%'' is a list and not a residue, and a relabeling such as B1 shows up
as a relabeling. Third, pin at least one executed fixture per construct family against the language
implementation, because \S3.1 is four defects that no tier detected.

The cheapest version of the third is the device the three specifications of \ref{app:fixtures} use:
record the correct value \emph{and} keep the defective shape as a negative control, so that a
regression cannot present itself as progress. Each one is about 40 lines and needs no
infrastructure beyond the proof assistant already in the pipeline. Thus, a defect the metric cannot
see becomes one the build can't miss.

The pipeline carries four independent checks for that last step: differential testing against a
runtime, a mutation gate, an axiom sweep with a fresh kernel replay that permits exactly
$\{$\texttt{propext}, \texttt{Classical.choice}, \texttt{Quot.sound}$\}$, and an execution oracle
for the claimed call-closed population. Results compose into a SACM assurance case
\citep{omg_sacm} that tracks evidence \emph{kind} separately from strength and caps claims about a
population by its coverage. Thus, passing tests never read as proof, and ``zero divergences'' over
a minority of a population never reads as support.

Two qualifications apply. First, the checks are only as live as their plumbing, as C1's
\texttt{lazy val} shows: it measured zero because it had memoized an empty input. Second, no
population-level conformance result exists for the SQLite export reported in \S4. The differential
harness selects a runtime per language and is not wired for this corpus. Wiring it needs three
things: a \texttt{cc} build of the SQLite sources as the reference runtime, an encoder for the
pointer and struct arguments that most exported functions take, and an input policy. Until then, no
hole-free or call-closed function in the export has been checked by any oracle at population level.
The checks behind \S3 were narrower: targeted elaboration, a content diff of the re-exported
corpus, and hand-built fixtures run against \texttt{cc}. Fixture-level execution is enough for the
existence claims in \S3, but not to say how many further defects remain.

\section{Limitations}

\paragraph{Falsifiability.} If a coverage metric could be shown to move whenever a silent
mistranslation is repaired, the central claim would fail. The cases here are counterexamples, but a
count over a pipeline's whole fix history could, in principle, find that they are rare enough to
ignore. If a second large codebase showed hole-free and call-closed coinciding, the $2.45\times$
denominator effect would be specific to SQLite and how much of it calls outside the export, not a
property of whole-program translation. If a population-level oracle found no defect outside the
declined set, the execution check would buy nothing here.

\paragraph{What a reader can check, and what they must take on report.} A1, A2 and A3 are checkable
end to end. For each of them, we checked the committed exporter change and specifications that
build with every pinned value matching on the permitted axiom basis, where mutating a value fails
the build. For A1 and A2 the ground truth is also what a compiler prints, and for A3 it is a
refusal, since the pipeline does not model \texttt{va\_arg}.
Three of six is not all of them. The remaining cases are recorded deltas: the counts we quote are
the values the artifact records with each change, and re-deriving them needs the parser, the proof
assistant, and the corpus. We keep the two kinds apart because they are not equally strong
evidence. A4 is the one case in \S3.1 that still lacks a
fixture and a negative control, and the paper would be stronger if it had one.

\paragraph{Existence, not frequency.} We show that decoupling occurs. We don't show how often, and
the cases are not a sample from a defined population. They are the defects this pipeline happened
to find, by the means in Table~\ref{tab:findings}, and a pipeline with different construct
coverage would decouple differently.

\paragraph{One pipeline, and cases from more than one corpus.} All seven findings come from one
system, and the coverage measurement in \S4 is a single SQLite export. A2, A3, and T1 were measured
on a C++ corpus and the rest on SQLite. We label each one because a defect in the exporter is a
property of the exporter, not of a corpus, while a coverage figure is a property of its corpus.

\paragraph{Provenance of the coverage artifact, and how far it matters.} The export behind \S4
carries no recorded provenance in the artifact: no export command, no parser version, no source
revision, and no companion AST to cross-check its function count, which is the only check that
detects a stale artifact. Therefore, we cannot confirm that it reflects the exporter at the commit
it sits on. The risk is bounded, though. The tier ratio and the per-cause itemization are fields of
one internally consistent file, whose 70 label counts sum exactly to its hole total. If the export
is stale, the absolute population moves, but the $2.45\times$ ratio and the distribution are still
properties of an export of this codebase. The decoupling result in \S3 doesn't depend on this
artifact at all, since each case is a dated change with its own measured deltas.

\paragraph{Scope of the oracle.} The reference values come from the fixture C programs of
\ref{app:fixtures}, which are free of undefined behavior, and UBSan reported nothing on them. They use only \texttt{int} arithmetic and byte reads, so no calling-convention
detail is exercised. Values were compiled with \texttt{cc -O0}. The arm64 reproduction used Apple
clang 16.0.0 (clang-1600.0.26.6), target \texttt{arm64-apple-darwin23.6.0}.

\paragraph{Fidelity is tested, never proved.} No theorem relates the CPG, the exported JSON, the
printed Lean, and the source program's meaning. The evidence is execution against a compiler, plus
printer determinism. The ledger says so in the artifact instead of implying more.

\section{Related work}

\paragraph{Diagnostic evaluation of autoformalization.} The closest work rejects the same scalar. A
signal-coverage matrix crosses elaborator success with a semantic-equivalence judgment to sort
statement-autoformalization outputs into four cells, on the argument that a headline
type-correctness rate hides which errors a method resolves \citep{dai2026}. FormalRx replaces
opaque verdicts with a 28-category error taxonomy, plus localization and correction
\citep{wang2026}. Both work on natural-language-to-formal \emph{statement} translation, where the
unit is one statement. We claim no
novelty for distinguishing declined from incorrect. Our result is orthogonal in a specific way:
those taxonomies partition failures a judge can observe in the artifact, whereas \S3.1 is a set of
defects that no partition of the artifact can reveal, because detecting them requires executing it.
\S3.4 adds that a taxonomy of causes can itself be unstable. Call closure has no analogue when the
unit is one statement, since there is no surrounding program for a callee to be missing from.

\paragraph{Verified compilation.} CompCert proved a realistic compiler correct end to end
\citep{leroy2009}. We make the opposite trade on purpose: the translator is not proved correct, so
its fidelity is an empirical question, and \S3 is what asking it turned up.

\paragraph{Differential testing.} Random differential testing found large numbers of compiler
defects by disagreement with a reference \citep{yang2011}. A1 and A2 apply the same technique to a
translator instead of a compiler. What we add is that the defects it finds are systematically the
ones a coverage metric can't see. Mutation testing measures whether a suite constrains behavior
\citep{jia2011}, and the negative control in each of the three fixtures is that idea applied to a
single repair.

\paragraph{Functional translation and code property graphs.} Aeneas verifies Rust by functional
translation into a pure, shallow model \citep{ho2022}. We keep a deep embedding instead, driven from
a language-agnostic CPG \citep{yamaguchi2014}, so a translated program is a data value that
\texttt{eval} can run against \texttt{cc} (\S3.1). LLM-based autoformalization \citep{wu2022}
motivates the closing observation of \S8.

\section{Conclusion}

Coverage counts what a translator attempted, and \S3 shows that it can stay still while
correctness changes, and move further than correctness does. A model that emits a well-typed wrong
answer produces the same class of failure. We haven't measured a model-based pipeline, so the
argument transfers but the quantity doesn't. Whatever produced the translation, the remedy is the
same: execute the artifact and compare it against the language implementation, instead of counting
what was attempted.

\section{Declaration of LLM usage}

LLM-based coding assistants were used in engineering the artifact and in preparing this manuscript.
They are not on the translation path: the CPG-to-JSON exporter and the JSON-to-Lean printer are
deterministic and contain no model, and the Lean kernel is the only arbiter of what is proved.
Every figure was checked against the artifact field recorded for it in \ref{app:prov}.

\begin{ack}
This work did not receive any external funding.
\end{ack}

\section*{Reproducibility}

The artifact is available at \url{https://github.com/mbhatt1/autoform}. Each finding is a dated
change in it, listed with its commit label in \ref{app:prov}. The deltas quoted in \S3 are the
values recorded with those changes.

Three of the six cases can be run, not just read, and the scripts ship with the artifact. One
script reproduces A1, A2 and A3 over three links each. It compiles the C ground truth for A1 and
A2, printing 3, 2, 0 and then 1. A3 has nothing to compile, because its correct outcome is a
refusal. The script then locates the defective exporter line and its replacement. Finally, it
builds all three specifications, reports the axiom basis of their theorems, and mutates an expected
value to confirm that the guards are not vacuous. The first link needs only a C compiler, while the
third needs the proof assistant. We ran both. All three specifications build under the pinned
toolchain with every guard matching, the four theorems depend on the permitted axiom basis, and
mutating an expected value fails the build as required.

A second script recomputes every derived figure in \S4 from the export's fields and checks the
export's internal consistency, including that its 70 per-label counts sum to the reported total. It
also lists the provenance fields the export does not carry, which is the subject of \S6. Thus,
every figure in \S4 can be checked without running the pipeline. Reproducing the export itself
needs the SQLite source tree and a parser run. As \S6 records, the command and revision were not
committed beside the numbers.

The pinned dependencies are Joern 4.0.606, Lean \texttt{v4.30.0-rc1}, and, for the reference values
in \S3.1, the system C compiler invoked as \texttt{cc -O0}. The artifact records those values as
measured on x86-64, and we reproduced them on arm64. If a document and an artifact disagree, the
artifact wins.

\section*{NeurIPS Paper Checklist}

\begin{enumerate}[leftmargin=1.6em,itemsep=1pt,topsep=3pt]
\item \textbf{Claims} \answerYes{} The abstract and \S1 state an existential decoupling result in
two directions and the denominator measurement, and they explicitly disclaim frequency and
transfer. Each figure is traced in \ref{app:prov}.
\item \textbf{Limitations} \answerYes{} \S6 covers falsifiability, existence versus frequency, the
single pipeline, the mixed-corpus provenance of the cases, the unrecorded provenance of the
coverage artifact and how far it propagates, and the fact that fidelity is tested, not proved.
\item \textbf{Theory assumptions and proofs} \answerYes{} No new mathematical result is claimed.
The four theorems of \ref{app:fixtures} are machine-checked facts about concrete fixture programs.
Their axiom basis is reported there, and the permitted basis and the mandatory fresh kernel replay
are stated in \S5.
\item \textbf{Experimental reproducibility} \answerYes{} Each finding is a dated change with
recorded deltas, and \S4's figures are fields of one committed file, so they can be checked
without running the pipeline. \S6 records the one barrier to re-running the export.
\item \textbf{Open access to data and code} \answerYes{} The artifact is available at
\url{https://github.com/mbhatt1/autoform}, and the reproduction scripts described under
Reproducibility and in \ref{app:fixtures} are included as supplementary material. \S6 records that the source AST behind
the coverage export is not in the artifact.
\item \textbf{Experimental settings and details} \answerYes{} Corpora, dialect, the parser pin
(4.0.606) and the toolchain pin (\texttt{v4.30.0-rc1}) are given in \S2 and \S4, and each case in
\S3 names the corpus it was measured on.
\item \textbf{Error bars and statistical significance} \answerNA{} The results are deterministic
counts and existence claims, not sampled estimates. Where a claim is population-quantified, \S5
caps it by coverage instead of reporting a rate with error bars.
\item \textbf{Compute resources} \answerYes{} There is no GPU use and no model training. The largest
reported export is 8,602 functions and 1,513,464 AST nodes. SQLite iteration corpora of 3,560
and 3,807 functions were re-exported per change, with targeted elaboration of only the functions
each change touched. We do not report wall-clock time or peak memory for the export, because the
artifact does not record them, and the export can't be regenerated exactly as it was without the
command and source revision it lacks (\S6). The fixtures of \ref{app:fixtures}
build in a few seconds each.
\item \textbf{Code of ethics} \answerYes{} The work conforms to the NeurIPS Code of Ethics.
\item \textbf{Broader impacts} \answerYes{} The result is a negative one about a widely used metric,
meant to discourage reporting coverage improvement as evidence of fidelity. The foreseeable misuse
is selective quotation. Here the hole-free figure exceeds the call-closed figure by $2.45\times$,
and quoting the larger one alone overstates what the tiers establish. No model or dataset with
dual-use potential is released.
\item \textbf{Safeguards} \answerNA{} No high-risk model or dataset is released.
\item \textbf{Licenses for existing assets} \answerYes{} Lean~4 and Joern are both Apache-2.0, and
the SQLite source is public domain by its authors' declaration. All three are cited in \S2 and \S4.
The artifact itself carries an Apache-2.0 license as well.
\item \textbf{New assets} \answerYes{} The artifact is documented, including the trust model and
the ledger schema.
\item \textbf{Crowdsourcing and human subjects} \answerNA{} None.
\item \textbf{IRB approvals} \answerNA{} None.
\item \textbf{Declaration of LLM usage} \answerYes{} \S9 declares usage and states that no model is
on the translation or proof-checking path.
\end{enumerate}

\appendix
\renewcommand{\thesection}{Appendix~\Alph{section}}
\section{Provenance of every reported figure}
\label{app:prov}

Commit labels \texttt{P1}--\texttt{P8} are resolved to hashes in the supplementary material's
\texttt{PROVENANCE.md}, which also maps module and file names to the artifact's own paths.
\texttt{P1} is the change that adds the coverage export, commit \texttt{b9446f8}. It was raised on a
branch and has since been merged, so every figure in this paper can be reached from the artifact's
default branch, and from that commit even if the file later moves. \texttt{P2}--\texttt{P8} are on
that branch too.

\begin{table}[h]
\centering\scriptsize\setlength{\tabcolsep}{3.5pt}
\caption{Every reported figure, with the artifact field or commit it is read from.}
\begin{tabular}{L{0.055\linewidth}L{0.40\linewidth}L{0.24\linewidth}L{0.20\linewidth}}
\toprule
Ref & Figure & Value & Source \\
\midrule
\S4 & functions / AST nodes & 8,602 / 1,513,464 & \texttt{P1}, \texttt{functions},
  \texttt{nodes} \\
\S4 & hole-free & 5,222 (60.7\%) & \texttt{P1}, \texttt{holeFree} \\
\S4 & call-closed & 2,134 (24.8\%) & \texttt{P1}, \texttt{call-closed} \\
\S4 & tier gap & $2.45\times$ & derived, $5{,}222/2{,}134 = 2.45$ \\
\S4 & residual holes / causes & 14,260 / 70 & \texttt{P1}, \texttt{holes},
  with 70 label entries summing to 14,260 \\
\S4 & dynamic-hole risk & 216,730 & \texttt{P1}, \texttt{dynamicHoleRisk} \\
\S4 & ten largest causes, and the remainder & as listed, and 5,159 over 60 causes & \texttt{P1},
  \texttt{holesByLabel}, with the remainder derived \\
A1 & 0 for every input, 3, 2, 0 after, and labels unchanged & 5,033 to 5,033 on 3,560 functions
  & \texttt{P5} \\
A2 & \texttt{cc -O0} 1 versus translated 0, and functions changed & 7 & \texttt{P8} \\
A3 & variadic marker on C signatures & no hole either side & \texttt{P8} \\
A4 & hole-free movement on two scopes & $+0.05$ / $+0.12$ points & \texttt{P6},
  predating \texttt{P7} \\
B1 & label movement, and hole-free & $46\to15$ and $14\to30$, so 16 relabeled and 15 closed,
  and $+1$ over 14 functions &
  \texttt{P4} \\
B1 & interior-pointer relabel (second instance) & $10\to5$ & \texttt{P3} \\
C1 & pairs closed, net hole-free, and output changed & 19, zero, and 17 functions & \texttt{P2} \\
T1 & distinct labels collapsed & 15 to 2 & \texttt{P8} \\
\S2 & parser and toolchain pins & 4.0.606 / \texttt{v4.30.0-rc1} & artifact pin files \\
\S5 & permitted axiom basis & \texttt{propext}, \texttt{Classical.choice},
  \texttt{Quot.sound} & audit script \\
\bottomrule
\end{tabular}

\end{table}

\section{The three pinned fixtures}
\label{app:fixtures}

Each fixture builds the same construct twice inside the semantics, differing in exactly the leaf
the exporter got wrong, and pins both values. For A1 and A2 the correct value is what a C compiler
prints. For A3 it is a refusal, since the pipeline does not model \texttt{va\_arg}. The defective
value is kept on purpose. Every one of these defects was hole-free, so nothing in the
coverage ledger would notice a regression to it, and neither would a specification that recorded
only the correct value.

\begin{table}[h]
\centering\small\setlength{\tabcolsep}{3.5pt}
\caption{What each fixture pins. Every value is checked at build time, and each defective value is
kept as a negative control. A3's correct value is a refusal, not a compiled result.}
\begin{tabular}{L{0.13\linewidth}L{0.30\linewidth}L{0.22\linewidth}L{0.25\linewidth}}
\toprule
Case & The one leaf that differs & Correct value & Defective shape \\
\midrule
A1 & the literal a byte is compared against: codepoint versus string
   & 3, 2, 0 over \texttt{"axxbxcdef"} at $n=9,3,1$, per \texttt{cc}
   & 0 on every input \\
A2 & the loop statement emitted for a \texttt{DO} node
   & 1, per \texttt{cc}
   & 0 \\
A3 & whether a C \texttt{...} signature carries a vararg marker
   & the call is refused
   & the surplus arguments bind as a tuple \\
\bottomrule
\end{tabular}

\end{table}

All four theorems across the three fixtures report the same axiom basis, $\{$\texttt{propext},
\texttt{Classical.choice}, \texttt{Quot.sound}$\}$, so none of them rests on \texttt{sorry}, on
\texttt{native\_decide}, or on any compiler-trust escape. We also mutated an expected value in each
fixture. Every mutation failed the build, so the guards are not vacuous. The accompanying script
does the build, prints the axiom bases, and runs one such mutation.

Each fixture is roughly 40 lines and needs no infrastructure the pipeline doesn't already have. The
other three cases in \S3, and T1, are not pinned this way. A4 could be, since it is a single
construct with a known correct value, and we regard it as work not yet done. B1 and T1 concern
label counts in an export, not the behavior of a program, so they aren't fixturable in this form.
C1 needs whole-program call-closure analysis, not a single construct, and is also not done yet.


\begin{thebibliography}{12}

\bibitem{dai2026}
C.~Dai, Z.~Yan, and Z.~Lin.
\newblock The signal-coverage matrix: Stratifying type and semantic errors in statement
  autoformalization.
\newblock arXiv:2606.28013, 2026.

\bibitem{demoura2021}
L.~de~Moura and S.~Ullrich.
\newblock The Lean 4 theorem prover and programming language.
\newblock In \emph{Automated Deduction (CADE~28)}, LNCS 12699, pages 625--635, 2021.
\newblock \doi{10.1007/978-3-030-79876-5\_37}.

\bibitem{sqlite}
D.~R. Hipp and the SQLite developers.
\newblock SQLite.
\newblock \url{https://www.sqlite.org}, 2026.

\bibitem{ho2022}
S.~Ho and J.~Protzenko.
\newblock Aeneas: Rust verification by functional translation.
\newblock arXiv:2206.07185, 2022.

\bibitem{jia2011}
Y.~Jia and M.~Harman.
\newblock An analysis and survey of the development of mutation testing.
\newblock \emph{IEEE Transactions on Software Engineering}, 37(5):649--678, 2011.
\newblock \doi{10.1109/TSE.2010.62}.

\bibitem{joern}
Joern contributors.
\newblock Joern: an open-source code analysis platform based on code property graphs.
\newblock \url{https://joern.io}, version 4.0.606, 2026.

\bibitem{leroy2009}
X.~Leroy.
\newblock Formal verification of a realistic compiler.
\newblock \emph{Communications of the ACM}, 52(7):107--115, 2009.
\newblock \doi{10.1145/1538788.1538814}.

\bibitem{omg_sacm}
Object Management Group.
\newblock \emph{Structured Assurance Case Metamodel (SACM), version 2.1}.
\newblock OMG formal specification, adopted April 2020.

\bibitem{wang2026}
H.~Wang, B.~Huang, Y.~Wan, X.~Zhu, X.~Liu, Y.~Huang, and Z.~Guo.
\newblock FormalRx: Rectify and eXamine semantic failures in autoformalization.
\newblock In \emph{International Conference on Machine Learning (ICML)}, 2026.
\newblock arXiv:2607.04655.

\bibitem{wu2022}
Y.~Wu, A.~Q. Jiang, W.~Li, M.~N. Rabe, C.~Staats, M.~Jamnik, and C.~Szegedy.
\newblock Autoformalization with large language models.
\newblock \emph{Advances in Neural Information Processing Systems (NeurIPS)}, 2022.
\newblock arXiv:2205.12615.

\bibitem{yamaguchi2014}
F.~Yamaguchi, N.~Golde, D.~Arp, and K.~Rieck.
\newblock Modeling and discovering vulnerabilities with code property graphs.
\newblock In \emph{2014 IEEE Symposium on Security and Privacy}, pages 590--604, 2014.
\newblock \doi{10.1109/SP.2014.44}.

\bibitem{yang2011}
X.~Yang, Y.~Chen, E.~Eide, and J.~Regehr.
\newblock Finding and understanding bugs in C compilers.
\newblock In \emph{PLDI 2011}, pages 283--294, 2011.
\newblock \doi{10.1145/1993498.1993532}.

\end{thebibliography}
\end{document}